\documentclass[twocolumn]{aastex61}
\usepackage{amsmath,amstext}
\usepackage[T1]{fontenc}
\usepackage{apjfonts} 
\usepackage[figure,figure*]{hypcap}
\usepackage{color, soul}
\usepackage{natbib}

\newcommand*{\rom}[1]{\expandafter\@slowromancap\romannumeral #1@}
\newcommand{\rnum}[1]{\uppercase\expandafter{\romannumeral #1\relax}}

\shorttitle{Transition Elements in Presolar SiC X Grains}
\shortauthors{Marhas and Sharda}

\begin{document}

\title{Transition Elements in supernova Presolar Grains: condensation vs. implantation}

\author{Kuljeet K. Marhas}
\affiliation{Planetary Sciences Division, Physical Research Laboratory, Ahmedabad, Gujarat 380009, India. Email: kkmarhas@prl.res.in}
\author{Piyush Sharda}
\affiliation{Dept. of Physics, Birla Institute of Technology and Science, Pilani, Rajasthan 333031, India}
\affiliation{Dept. of Electrical and Electronics Engineering, Birla Institute of Technology and Science, Pilani, Rajasthan 333031, India}

\begin{abstract}
We compute the concentrations of five transition elements (Cr, Fe, Co, Ni and Zn) via condensation and implantation in supernova presolar grains (Silicon Carbide Type X) from the time they condense till the end of free expansion (or pre-Sedov) phase. We consider relative velocities of these elements with respect to grains as they condense and evolve at temperatures $\le$ 2000 K, use zonal nucleosynthesis yields for three core collapse supernovae models - 15 M\textsubscript{\(\odot\)}, 20 M\textsubscript{\(\odot\)} and 25 M\textsubscript{\(\odot\)} and an ion target simulator SDTrimSP to model their implantation onto the grains. Simulations from SDTrimSP show that maximal implantation in the core of the grain is possible, contrary to previous studies. Among the available models, we find that the 15 M\textsubscript{\(\odot\)} model best explains the measured concentrations of SiC X grains obtained from Murchison meteorite. For grains where measured concentrations of Fe and Ni are $\ga$ 300 ppm, we find implantation fraction to be $\la$ 0.25 for most probable differential zonal velocities in this phase which implies that condensation is dominant than implantation. We show that radioactive corrections and mixing from the innermost Ni and Si zones is required to explain the excess Ni (condensed as well as implanted) in these grains. This mixing also explains the relative abundances of Co and Ni with respect to Fe simultaneously. The model developed can be used to predict concentrations of all other elements in various presolar grains condensed in supernova ejecta and compared with measured concentrations in grains found in meteorites.
\end{abstract}


 \section{Introduction}
\label{s:intro}
Supernovae (SN) are rare astronomical events (\textasciitilde3 per century) which provide an enormous wealth of information on stellar evolution and nucleosynthesis of elements. Studying these singular events has been quite challenging, given the large scale of energies (10$^{51}$ J), mass (\textasciitilde10-200 M\textsubscript{\(\odot\)}) and temperatures (\textasciitilde10$^9$ K) involved in supernovae physics. As the matter moves outwards after an SN explosion, expanding envelopes of stellar ejecta cool adiabatically. Eventually, the condensation of solid grains of sizes ranging from nanometers to a few tens of microns \citep{1982AdSpR...2...13G,1988mess.book..984N} takes place. These grains which are found in meteorites are termed as presolar grains, owing to their time of formation which precedes that of the solar system \citep{1987Natur.326..160L,1990Natur.345..238A}. At times, these are also referred as circumstellar grains and the ones condensing around supernovae have been termed X grains \citep{1990Natur.345..238A,1992ApJ...394L..43A} or SUNOCONS \citep{1975ApJ...199..765C,1975ApJ...198..151C,2002ApJ...578L..83C}. These grains not only carry imprints of the nucleosynthesis environment within the star but also provide an insight into post explosion ejecta evolution and related processes that go on in the supernovae environment (SNe). 

Of the various minerals (like nitrides, oxides, carbides) condensing in the SNe, largely, Silicon Carbide (SiC) has been studied more in the laboratory for its morphology, along with elemental and isotopic compositions \citep{1998AREPS..26..147Z,2010ApJ...719.1370H,2014LPICo1800.5051H,2017LPI....48.2331L}. The SiC X grains from supernovae constitute \textasciitilde1\% of the total presolar SiC grains and are characterized by higher ${^{12}\textnormal{C}}/{^{13}\textnormal{C}}$ and lower ${^{14}\textnormal{N}}/{^{15}\textnormal{N}}$ ratios than solar abundances \citep{2005ChEG...65...93L}. Infact, few of them contain high ${^{26}\textnormal{Al}}/{^{27}\textnormal{Al}}$ ratios (upto 0.6) and all of them are consistently endowed with $^{28}$Si enrichment. Supernovae genesis of these grains is strongly supported by traces of $^{44}$Ti and $^{49}$V, which are produced during the explosive nucleosynthesis of $^{44}$Ca and $^{49}$Ti respectively \citep{1996ApJ...462L..31N,2002ApJ...576L..69H,2014AIPC.1594..307A}, suggesting these grains condense around type II core collapse supernovae (CCSN) \citep{1992ApJ...394L..43A}, although the possibility of finding X type SiC grains in SN Ia cannot be ruled out \citep{1997ApJ...486..824C,1998M&PSA..33R..10A}. 

The pre-explosion SN structure can be associated with chemically distinct zones marked by the most abundant element(s), \textit{viz.}, $^{56}$Ni, $^{28}$Si, $^{20}$Ne, $^{16}$O, $^{12}$C, $^{4}$He and $^{1}$H, in the order of successive hydrostatic burning stages \citep{1995Metic..30..325M}. The inner most shells are rich in transition elements since high binding energy of Fe stops further fusion. Post explosion, neutron sources (like, some radioactive elements) and most abundant elements in each of these zones transcend into regions with variable abundances of certain isotopes which are the diagnostic signature of a supernova.  Extensive mixing occurs due to Rayleigh-Taylor (RT) instabilities at the zonal boundaries due to which material from the interior is mixed with the outermost envelopes, as predicted by simulations \citep{2003A&A...408..621K} and confirmed by laboratory measurements  \citep{2002ApJ...564..896D,2009ApJ...696..749K}. Similarly, H from the outermost zone gets mixed with He and other zones in the interior in a series of reverse mixing stages because of Richtmyer-Meshkov instabilities \citep{1992ApJ...400..222B,1999ApJ...511..335K}. Isotopic ratios of elements can provide information on the nucleosynthesis pathways, evolution and mixing occurring at the peripheries of various zones.

While \cite{1998Sci...281.1165V} proposed elemental abundances to be highly sensitive to grain size as evidence for ion implantation, \cite{2006M&PSA..41.5248M} proposed the existence of a negative correlation between elemental concentration and grain size. To understand and relate the isotopic signatures observed in presolar grains with post supernovae RT mixing, we have constructed a model to simulate trace ion implantation in the SNe. Specifically, the simulations study high energy Cr, Fe, Co, Ni and Zn from SN explosion interacting with 1 $\micron$ and 5 $\micron$ SiC grains. The model proposed can be applied to all isotopes eventually; we choose to begin with transition elements because their implantation has not been studied in detail, unlike the implantation of noble gases \citep{2003M&PSA..38.5200V,2004ApJ...607..611V,2008M&PS...43.1811H} and low mass elements like Li and B \citep{2001ApJ...551..478H,2007M&PS...42..373L}. In fact, some predictions for the rare earth elements have also been made, see, for e.g., \cite{2006ApJ...647..676Y}. The idea is to de-alienate ion implantation with direct condensation and compare with observations from laboratory studies. Incorporating RT mixing and differential zonal velocities make the model highly versatile because one can vary these parameters to correspond to measured concentrations and the resulting parameter set can give a handle on physical conditions present in the grain surroundings at the time of implantation, not to mention the mass of the progenitor in the first place where these grains must have condensed. 

We study backscattering, implantation, transmission and sputtering (hereafter, BITS) processes to check the total implantation of these isotopes in a spherical SiC grain. In section \ref{s:sec2}, we talk about supernovae nucleosynthesis (which leads to the production of heavy ions in question) and condensation of SiC in SNe. Section \ref{s:implan} discusses the theory of transition ion implantation in presolar grains and section \ref{s:sdtrimsp} describes how the ion target simulator SDTrimSP is set up. We present all the results and calculations in section \ref{s:disc} - section \ref{s:disc1} gives a summary of BITS processes and section \ref{s:disc2} lays down all the calculations by taking up the example of Cr. In the last subsection (\ref{s:disc3}), we compare our calculated concentrations with the ones measured in the laboratory, improve them by adding appropriate corrections and discuss reasons for similarities and discrepancies. Finally, we summarize our results in section \ref{s:summ}. In Appendix \ref{s:append}, we show the derivation of model used for studying ion transmission through the grain. 

\section{NUCLEOSYNTHESIS AND CONDENSATION WITHIN SNe}
\label{s:sec2}
Nucleosynthesis in CCSN has been studied in detail over the last couple of decades after major breakthroughs in computational astrophysics. Although 3D hydrodynamic models have also been developed \citep{2015A&A...577A..48W,2015A&A...581A..40U,2017MNRAS.472..491M}, they are still subject to scientific scrutiny as the explosion mechanism is not properly understood. To explore the implantation of trace elements into presolar grains, it is vital to analyze their evolution after explosion. We use the zonal nucleosynthesis yields provided by one of the hydrodynamic models, along with the surrounding conditions to predict the amount of transition ion implantation in grains condensed in the ejecta. 

There are numerous models which lay down nucleosynthesis yields from CCSN explosions \citep{1957RvMP...29..547B,1986ARA&A..24..205W,1995ApJS..101..181W,1996ApJ...460..408T,2002ApJ...567..532H,2002ApJ...576..323R,2006NuPhA.777..424N,2006ApJ...653.1145K,2007PhR...442..269W,2010ApJ...724..341H,2015ApJ...808L..43P,2016ApJ...821...38S}. In this work, we utilize the zonal yield sets from \cite{2016ApJ...821...38S} (hereafter, S16), which used the modified 1D hydrodynamic code KEPLER\footnote{https://2sn.org/kepler/doc/Introduction.html} along with P-HOTB. P-HOTB stands for Prometheus-Hot Bubble and was used to study core collapse \citep{1996A&A...306..167J,2003A&A...408..621K} whereas KEPLER was used to evolve the star along zero age main sequence (ZAMS) and calculate nucleosynthesis yields and light curves \citep{1978ApJ...225.1021W}. For models which exploded, isotopic yields were generated post explosion. The zonal yields were obtained for three particular models - 15.2 M\textsubscript{\(\odot\)}, 20.1 M\textsubscript{\(\odot\)} and 25.2 M\textsubscript{\(\odot\)} \textasciitilde200 seconds after the explosion, before any mixing could take place\footnote{T. Sukhbold, \textit{Private Communication}}. Although all models in S16 dataset assume solar metallicity and do not take into account the effects of rotation, they can account for: 1.) detailed neutrino transport calculations using an improved explosion mechanism as compared to \cite{2002ApJ...576..323R} and \cite{2007PhR...442..269W}; 2.) a central engine which considers matter inside the collapsed core unlike certain other models which investigated only the matter exterior to the central engines used and; 3.) unlike previous nucleosynthesis models, all models\footnote{Each model has a particular progenitor mass.} used here are not exploded by injecting artificial energy because: a.) models below 15 M\textsubscript{\(\odot\)} almost always explode, b.) models in 20-30 M\textsubscript{\(\odot\)} rarely explode and c.) most models above 30 M\textsubscript{\(\odot\)} implode and become black holes (see Figure 14 in S16 for the probability of explosion of different progenitor masses). Infact, the few models above 30 M\textsubscript{\(\odot\)} in which explosion does take place is due to their core being ripped apart by winds to sizes comparable to \textasciitilde15 M\textsubscript{\(\odot\)}. The decimals in these models might seem bizarre; the reason is that the authors have tried to explode all possible progenitor masses in steps of 0.1 M\textsubscript{\(\odot\)} between 12-30 M\textsubscript{\(\odot\)}, however, 15.0, 15.1, 20.0, 25.0 and 25.1 M\textsubscript{\(\odot\)} imploded in their simulations. This apparently small change in progenitor mass which leads to an altogether different end scenario is due to small but significant variations in the progenitor compactness \citep{2011ApJ...730...70O} rather than the central engine characteristics \citep{2015ApJ...801...90P}. This effect is more pronounced near progenitor masses of $\sim$20 M\textsubscript{\(\odot\)} because the carbon burning stage changes to the radiative pathway from a convective mechanism. Infact, it has been recently shown that two similar progenitors with identical masses but slightly different input physics can lead to totally different scenarios \citep{2017arXiv171003243S}. Thus, it is not unusual for such stark differences to show up between two similar progenitor stars. Throughout this paper, we frequently approximate 15.2 to 15, 20.1 to 20 and 25.2 to 25 M\textsubscript{\(\odot\)} models for the sake of simplicity.

\begin{figure*}
\figurenum{1}
\centering
\label{fig:fig1}
\includegraphics[width=7in, height=5in]{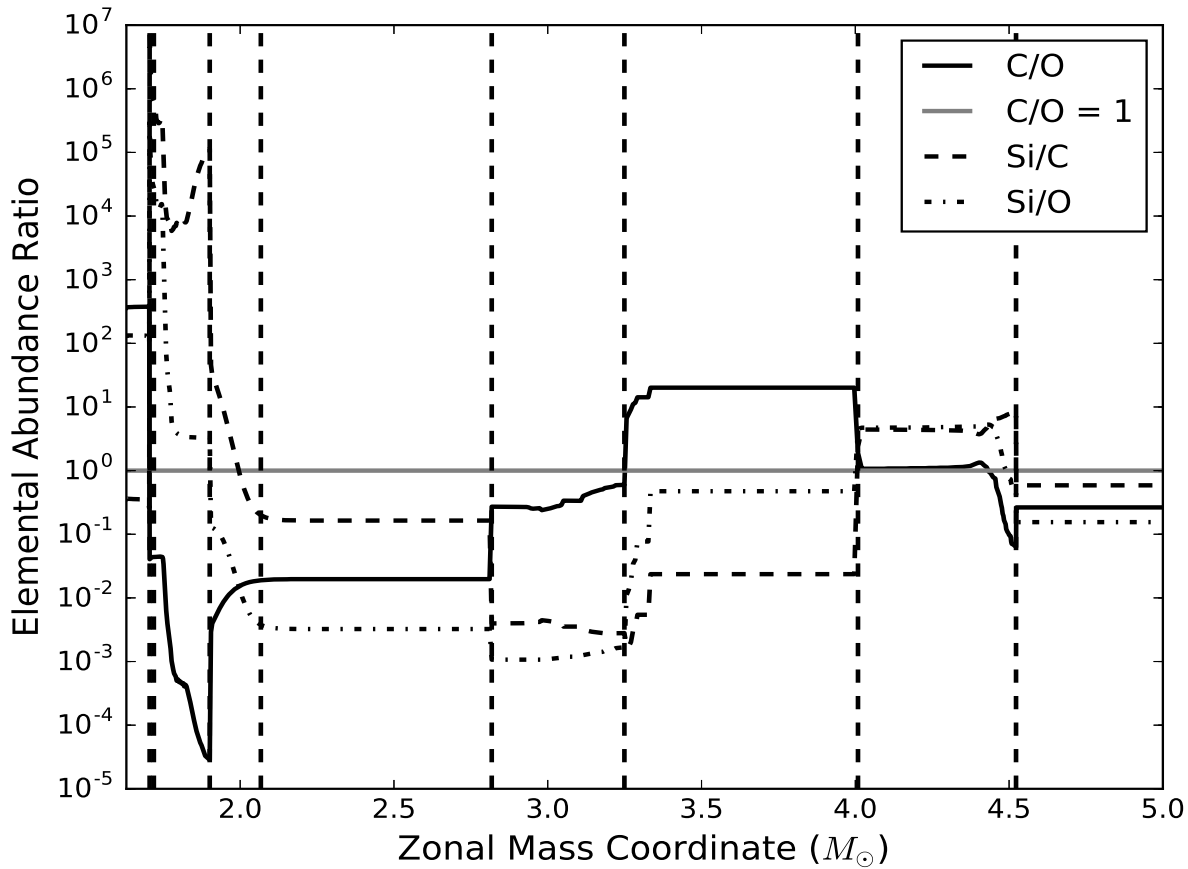}
\caption{Semi-logarithmic plot showing the variation of C/O, Si/C and Si/O with zonal mass for a 15.2M\textsubscript{\(\odot\)} star. We also show a reference line for C/O = 1. Vertical dotted lines mark zones classified by most abundant elements: Ni, Si/S, O/Si, O/Ne, O/C, He/C, He/N and H respectively. C/O >1 in the He/C and He/N zones, which have the highest probability of SiC formation. This ratio sharply decreases to \textasciitilde0.8 and then to \textasciitilde0.014 in the inner regions but remains essentially constant for the outer H and He envelopes. We only plot the figure till a zonal mass coordinate of 5.0, after which the ratios remain constant throughout the H zone.}
\label{fig:one}
\end{figure*}

SiC condensation could have taken place either in the inner shell or in the outer He shells where zonal C/O > 1 \citep{1979GeCoA..43.1455L,1997AIPC..402..391L}. The inner shell where $^{4}$He is thought to be the most abundant isotope is negligible in size as compared to its neighboring shells, for the three models in consideration. In fact, we only observe a clear $^{4}$He dominated shell in the 25.2 M\textsubscript{\(\odot\)} explosion model. Nevertheless, in the inner region, SiC condensation can be more prevalent in large progenitor mass models (like the 25.2 M\textsubscript{\(\odot\)} model) having more energetic explosions, provided the temperature is brought down to \textasciitilde2000 K within a few hundreds of days because at T > 2000 K, SiC is assumed to be stable only in its molecular form and does not condense to form a grain \citep{2013ApJ...776..107S}. The SiC grains formed in the inner region (if any) are, however, prone to destruction due to excessive sputtering by ions present in the nearby O rich regions or due to scattering by high energy particles. Moreover, TiC and SiS have been found to be more stable condensates in the inner regions where C/O <1, rather than Graphite or SiC \citep{2001GeCoA..65..469E,2013ApJ...776..107S}. Thus, we do not study SiC condensation in the inner regions and focus only on the outer He rich zone. However, \cite{1999Sci...283.1290C} have argued against the limiting criterion of C/O >1 by proposing the formation of large (up to few $\micron$) C grains in the ejecta interior if CO molecules are destroyed by radioactive nuclei early on and C atoms are freed. Notwithstanding such evidence for the formation of SiC, we have considered SiC condensation in an environment where C/O >1. These regions can be identified in the mass fraction plot against the mass of the star, as shown in Figure \ref{fig:fig1} for a 15.2 M\textsubscript{\(\odot\)} star to be between 3.2-4.0 M\textsubscript{\(\odot\)}. Similarly, we obtain the C/O >1 regions as 5.0-6.1 M\textsubscript{\(\odot\)} and 7.2-8.2 M\textsubscript{\(\odot\)} for 20.1 M\textsubscript{\(\odot\)} and 25.2 M\textsubscript{\(\odot\)} models respectively. Typically quoted condensation temperatures for SiC in these zone are $\le$ 2000 K (\citealt{1978ApJ...219..230L,1978M&P....19..169L,1995Metic..30..661L}), which are similar to those predicted for the solar neighborhood inspite of the lower pressure present in SNe, due to the depletion of Hydrogen (except for the outer H shell, see \citealt{2013M&PSA..76.5121G}). 

Observations of expanding envelopes of supernova ejecta suggest grain condensation can occur as early as 300-600 days \citep{1989ApJ...344..325K,1997AIPC..402..317W,2003ApJ...598..785N}. A recent work by Stephen et. al. (GeCoA 2017, in press) highlights a similar condensation time for Sr, Zr and Ba isotopes found in presolar SiC X grains. On the other hand, a much delayed formation (later than 1100 days) is proposed by \cite{2015A&A...575A..95S} in the outer region after explosion for a 15 M\textsubscript{\(\odot\)} stellar model, owing to 1.) the presence of He$^+$ (the presence of which will not let temperature decrease below 2000 K, see also \citealt{1990ApJ...358..262L}) for the first 1000 days after explosion and 2.) more efficient rates of condensation for carbon dust than SiC \citep{2009ApJ...703..642C}. Further, they also predict that SiC condensation starts as late as \textasciitilde1740 days, for a homogeneous ejecta model for a 19 M\textsubscript{\(\odot\)} star. An earlier condensation (\textasciitilde900 days) can be achieved if one takes into account the clumpy model, which says that the ejecta is no longer homogeneous after a few hundred days and is separated into spherical clumps owing to the finger like projections generated due to RT instabilities \citep{1992ApJ...392..118C}. 

After explosion, the immediate major source of heat present is in the form of high energy radioactive elements and condensation can start quite early in regions where such materials are in scarcity. In the inner regions, subsequent presence of $\gamma$ rays and Compton electrons \citep{0004-637X-562-1-480,0004-637X-638-1-234} is detected which ionize all material they encounter and hence, the ejecta attains very high temperatures (\textasciitilde10$^6$ K or more). Additionally, the presence of UV radiation due to the degradation of $\gamma$ rays can also cause the destruction of grains, although its effect is not as pronounced as compared to that of Compton electrons \citep{2009ApJ...703..642C}. The reverse shock while traveling inward also tends to heat the ejecta and reaches the remnant core by the time free expansion phase (also called the pre-Sedov phase, \citealt{1990ApJ...356..549S})\footnote{The term 'free expansion phase' refers to the initial few hundred years after a SN explosion, where the ejecta moves outwards with negligible deceleration. It is also called the pre-Sedov phase. This term is not really accurate because a reverse shock already starts traveling inwards during this phase which can heat the material it encounters to X-ray emitting temperatures.} ends. Its effect on SiC condensation is not clearly understood. 

C$\,$>$\,$O could also lead to some of the C being tied up as CO, thus delaying the formation of SiC; unless radioactive ions like $^{56}$Co can dissociate CO and free C atoms over a timescale of months, as suggested by \cite{0004-637X-562-1-480,2013ApJ...769...38Y}. However, the radioactive ions carry heat with them; thus, their presence could increase the temperatures to more than 2000 K \citep{1994ApJS...92..527N,0004-637X-842-1-13}. In O rich regions, SiO formation is more efficient and thus preferable than SiC, however, its total mass strongly decreases from its initial value at 200 days to that at 1500 days \citep{2013ApJ...776..107S}. It is also believed that SiO is the starting molecule in the formation of SiC. This is in direct support of an often quoted larger value of Si/C, since if Si>>C, SiC condensation can be straightforward because even after some Si atoms being locked in by O in the form of SiO, enough Si can be present to get condensed to SiC. A recent study by \citet{2017ApJ...843...57D} shows a direct dominance of radiative formation of SiC over SiO in the outer He rich zones. This is also confirmed by ALMA observations of SN1987A \citep{2017MNRAS.469.3347M} where the authors find only 10\% of the total Si synthesized in SNe to be locked up in SiO. Additionally, the authors also report a deficiency of SiS as compared to theoretical models \citep{2013ApJ...776..107S} whose over production could decrease the availability of Si for SiC. 

Overall, condensation of SiC primarily depends on temperature of the ejecta and concentration of the key species. The grain formation rate first increases as the temperature cools down, and then it decreases as the concentration of key species decreases. To demonstrate our calculations, we consider SiC condensation as late as 1700 days after explosion, in line with the recently proposed kinetic model of SiC formation by \cite{2017ApJ...843...57D}. By this time, the temperatures in the He zones are in the order of a few hundred K \citep{2013ApJ...776..107S}. We call this delayed condensation because other grain species are believed to undergo condensation at earlier times \citep{2001MNRAS.325..726T,2003ApJ...598..785N,2015A&A...575A..95S}. We urge the reader to go through \cite{2013M&PSA..76.5121G} and 
\cite{2013ApJ...776..107S} for a full zonal sequence of condensates produced in the supernova ejecta. 

\section{ION IMPLANTATION: TRANSITION ELEMENTS IN 
SiC GRAINS}
\label{s:implan}
\cite{2003ApJ...594..312D} have described the reverse shock mechanisms which propel the grain outwards in the ejecta. This model provides grain velocities of the order of a few hundreds of km s$^{-1}$ during the first few years, wherein they take the grain velocity to be 60\% that of the shock velocity. Similarly, \cite{2006ApJ...648..435N} suggest a value of 75\%. This is due to the deceleration of ejecta when it collides with reverse shock and formation of a contact discontinuity \citep{1999ApJS..120..299T}. Considering these values, we have carried our simulations for ion velocities in the range 1000-6000 $\mathrm{km}\,\mathrm{s}^{-1}$. 

We focus on these grain sizes because such grains are most probable to survive sputtering in the SNe (and later on in the interstellar medium (ISM)). Grains smaller than these sizes are highly prone to destruction within the SNe itself \citep{1994ApJ...433..797J,1996ApJ...469..740J}. Additionally, high velocity implantation (we study in this work) will be lesser in smaller grains since more ions will be able to transmit through them. Infact, it has been shown that grains smaller than \textasciitilde0.05 $\micron$ are destroyed due to excessive sputtering whereas \textasciitilde0.05-0.2 $\micron$ sized grains are trapped into a dense shell, making it impossible for them to be ejected into the ISM \citep{1995GeCoA..59.1411A,2007ApJ...666..955N}. Moreover, $\micron$ sized SiC grains condensed in SNe have been proposed to survive the SN shocks and get ejected into ambient ISM \citep{2004ApJ...614..796S}. Their longer lifetime as compared to graphite is another reason put forth for explaining SiC in $\micron$ sized presolar grains \citep{2016P&SS..133...17H}. Some smaller grains are also believed to have accumulated to form a larger grain size which can ensure their survival, specially due to charge separation between the smaller and larger grain and subsequent coagulation \citep{1990ApJ...361..155H}. Majority of SiC grains chemically separated in laboratory have a size of \textasciitilde0.1-10 $\micron$. Lastly, we note that our choice of grain sizes is based on the comparison of our calculated concentrations with those measured through NanoSIMS by \cite{2008ApJ...689..622M} where the grain sizes were of the order of a few $\micron$. Though grain size as large as 50 $\micron$ has also been found in meteorites which is believed to originate from red giant stars (e.g., \citealp{2009PASA...26..278G}), SN grains have been reported to have smaller diameters, see, for e.g., \citealt{2013GeCoA.120..628A,2016ApJ...820..140L}. The largest supernova SiC grain found till date is the famous Bonanza grain, with a size of roughly 30 $\micron$ \citep{2011LPI....42.1070Z}.

Isotopes of Cr, Fe, Co, Ni and Zn relevant to this study are all created during the explosive burning of Si. A material when heated to 5 billion K experiences nuclear statistical equilibrium (NSE) \citep{1994Metic..29Q.503M}. For a 15.2 M\textsubscript{\(\odot\)} star, this temperature is achieved in a radius of around 3670 km, which encloses about 2.02 M\textsubscript{\(\odot\)} (S16). In general, Cr is one of the products of Si burning whereas Co, Ni and Zn form during the alpha rich freeze out phase. Si burns via a series of photo disintegration reactions, producing alpha particles, which then react with the quasi equilibrium group (QSE) above $^{28}$Si to form Fe group elements \citep{2017arXiv170106786W}. During explosive burning, the nuclear burning timescale and the hydrodynamic time scales eventually become comparable, so the ejecta cools and expands before the alpha particles released from initial photo disintegration manage to get captured (thus called alpha-rich freeze out). These alpha particles eventually get assembled to heavier nuclei on a hydrodynamic timescale, to produce elements like Ni and Zn \citep{2002abcd...74.1015S} within seconds of the explosion\footnote{2D simulations of core collapse supernova find a higher yield of Zn as compared to S16, see \citealt{2005ApJ...623..325P,2017arXiv170106786W}}. After 2.02 M\textsubscript{\(\odot\)}, the O rich shell is present, where at temperatures ranging from 3-4 billion K, elements like Ca are produced. Fe is mostly produced by SNe Ia\footnote{Fe yields given in S16 models are calibrated to their upper bounds calculated in P-HOTB in S16 since it is underproduced in massive stars}.

All these ions which are produced in the innermost shells travel with velocities of the order of a few thousands of $\mathrm{km}\,\mathrm{s}^{-1}$. Grains condensing around the stellar envelopes provide a surface to stick on for these ions - analogous to cool balls in the middle of hot gas (A. Sarangi, \textit{private communication}). Since the grain area is quite larger than the size of particles in the gas, the rate of radiation of the grains is very large as compared to the gas particles. We discuss this transport (mixing) in detail in section \ref{s:disc3}.

\section{ION TARGET SIMULATOR SETUP}
\label{s:sdtrimsp}
TRIM (Transport of Ions in Matter) is a program in the SRIM (Stopping and Ranges of Ions in Matter) package, developed by \cite{2002srim...74.1015Z,2010NIMPB.268.1818Z}. TRIM was primarily used for studying implantation and backscattering in the field of materials science. It is now known that sputtering yields generated from SDTrimSP (developed by \cite{2011srim...74.1015M}; SD stands for Static-Dynamic, reflecting the fact that SDTrimSP can also work with dynamic targets, where composition of target changes as more ions are incident on it) better fits the experimental simulations than TRIM. SDTrimSP also offers a wide range of choices of input parameters like target temperature, choice of interaction potential and multiple bombardment of ions with varying velocities. A significant difference between TRIM and SDTrimSP arises because TRIM does not take into account the inelastic energy losses of the ions. We report calculations using implantation fraction obtained from SDTrimSP (Version 5.07). We obtain implantation data for relative velocities ranging from 1000-6000 $\mathrm{km}\,\mathrm{s}^{-1}$ and perform a total of 6400 iterations per ion. This choice is motivated by 1.) Vikram100 HPC processing capabilities and 2.) negligible change in statistics for iterations $\ge$ 3200.

TRIM and SDTrimSP only work for planar targets, so we extend their results to a spherical target to analyze SiC grains because it is the simplest structure we can assume the grains possess and spherical surfaces tend to be the most stable due to least surface area. The approximation we use also accounts for grain irradiation from all directions. 

Out of BITS processes, backscattering does not need a different geometrical model since the ion is not interacting with the grain's interior. Also, backscattered ions interact with the grain's outer surface for a very short time, causing only surface erosion, which can be neglected when compared with sputtering due to other incident ions. For spherical modeling of implantation profiles, we follow the method developed by \cite{vyvsinka2009depth}. This model gives different weights to ions incident at various angles getting implanted into the grain at diverse depths by considering the sphere as a regular polyhedron whose number of sides are decided by the number of bins of incident angles used. The weight is thus a product of the depth profile and surface area of the cross section of the grain for different angles (see Figure 2 in their paper). 

For spherical modeling of transmission and sputtering profiles, we follow a different approach, whose framework and calculations are derived in Appendix \ref{s:append}. We develop it to be coherent with the content of output data files generated in SDTrimSP. The output file provides the final position of the ion before leaving the target. Taking x axis to be the reference, we treat the distances covered in y and z directions separately and apply the model described in Appendix \ref{s:append} on both the axes. This model assumes that ions travel in a straight path (not necessarily parallel to the x axis) inside the grain till it is ejected out. The weights are the ratio of extra length traversed in planar targets to with the total length traversed in planar targets. This extra length is the distance the ions would not be traveling if the target was spherical, since they would have been transmitted at a shorter distance. The model takes as input the last known coordinate of the ion near the grain surface and the angle at the time of ejection. Then, assuming a straight line trajectory, it backtraces the ion to its place of origin (entrance). Based on the extra distance the ion had to travel for planar surface and the decrement in its kinetic energy inside the grain, for each ion (in the output file), the number of ions that could have been transmitted had the surface been a sphere is calculated. The straight line approximation is valid in the velocity interval where transmission dominates, since at these velocities, the ion cruises through the grain in an approximately linear trajectory. We confirm this by tracking ion trajectories throughout the target and find them to be approximately linear. This provides reasonably good results, considering the weights are applied to each ejected species individually, which is different from the model described for implantation, where weights are applied in chunks decided by the bin size. Following the notations given in section \ref{s:append}, we note that $L_p/L_{sph}$ >1, where $L_p$ is the length on planar surface and $L_{sph}$ is the projected length traversed in the sphere. We assume that transmission in 1D is inversely proportional to the distance traveled and is co-dependent on energy of the incident ion, since an ion with relatively less kinetic energy may also get transmitted if the angle of incidence is high.

Sputtering yields are highly sensitive to surface binding energies (hereafter, SBE) and lattice binding energies. It is a general practice to use heat of sublimation as SBE, however, the results do not match the experimental data obtained (at energies quite lower than the one discussed in this paper) specially for strong electronegative elements like O, C, etc. since strong ionic bonds can form between atoms in the top layer and those in the bulk \citep{WITTMAACK201237,MUTZKE2008872}. We use the model developed by \cite{2005ApSS..239..273K}, which takes into account weighted contributions from ionic and covalent bonds to calculate SBE. We provide all analysis without considering relativistic effects which are negligible at the range of velocities in question. In subsequent sections, we frequently use the terms 'larger' and 'smaller' grain to refer to 5 and 1$\micron$ grains respectively.

\section{DISCUSSION}
\label{s:disc}
\subsection{BITS Processes\footnote{All the discussion in this subsection is with respect to a 1$\micron$ grain, unless stated otherwise. }}
\label{s:disc1}
Using information from typically quoted shock and ion velocities in young supernovae (for e.g., \citealp{1987ApJ...315L.135K,2005ApJ...619..839C,diehl2013astrophysics}), we use an upper bound of 6000 $\mathrm{km}\,\mathrm{s}^{-1}$ for transition elements' ion implantation in SiC. At relative velocities higher than 6000 $\mathrm{km}\,\mathrm{s}^{-1}$ for these elements, \textasciitilde99.8\% of ions get transmitted through the grain. The remaining \textasciitilde0.2\% which get implanted have a major contribution from extremely oblique angles of incidences, which are mathematically possible but physically rare as compared to equatorial ion bombardments. At relative velocities lower than 1000 $\mathrm{km}\,\mathrm{s}^{-1}$, most ions get backscattered or are implanted into the upper surfaces of grains which have a high probability of being lost due to sputtering and erosion, etc. We observe that the contribution of extremely oblique angles (> 75$^{\circ}$) in implantation fraction is quite low ($\la$ 1\% for velocities < 3000 $\mathrm{km}\,\mathrm{s}^{-1}$ and $\la$ 0.2\% for velocities > 3000 $\mathrm{km}\,\mathrm{s}^{-1}$). The fraction of ions backscattered decreases as incident velocity is increased to 6000 $\mathrm{km}\,\mathrm{s}^{-1}$. Backscattered ions incident at oblique angles undergo a maximum loss of \textasciitilde36\% in kinetic energy while the ions backscattered at around 45$^{\circ}$ face a maximum reduction of only 7\%. Most runs did not give any backscattering for normal angles.

The low sputtering yields of Si and C by transition elements with velocities outside this range eliminate any possibilities of them significantly impacting the grain. At 6000 $\mathrm{km}\,\mathrm{s}^{-1}$, we find a sputtering yield of \textasciitilde0.06 for C and \textasciitilde0.08 for Si per incident transition ion, which is negligible as compared to the ones obtained in the lower energy range (\textasciitilde2-10 per incident ion). Maximum sputtering is observed at oblique incident angles, since at such angles more momentum transfer can take place and more atoms from the surface can be easily knocked out. Maximum damage to the grain is caused in the range of velocities showing maximum interaction (or equivalently, implantation). Sputtering yields can go as high as \textasciitilde28 times while the angle of incidence is changed from 0$^{\circ}$ to 85$^{\circ}$. For example, taking a numerical value of 20 ions sputtered per incident ion for Si (as observed in a run for Cr ions with a relative velocity of 1000 $\mathrm{km}\,\mathrm{s}^{-1}$ at 85$^{\circ}$ angle of incidence at 300 K), we see that a total of 10$^{-8}$\% of Si ions have been knocked out from the grain (assuming the process goes on for a few hundred years), which is still low to cause any significant changes. Also, for velocities near 1000 $\mathrm{km}\,\mathrm{s}^{-1}$, the sputtering yields were more at higher temperatures than at room temperature, which is a straightforward consequence of more atoms being knocked out by hotter incident ions. C has a higher sputtering yield than Si at T $\le$ 800 K and vice-versa, irrespective of the angle of incidence, which can be attributed to the fact that lighter atoms have a higher cross section to interact with the collision cascade and are thus more easily sputtered out (see review by \cite{SMENTKOWSKI20001} and references therein).	 

We note that the sputtering effects obtained from simulations only consider sputtering by a particular ion incident on the grain and associated collision cascades, whereas in reality there can be other fast moving ions hitting the grain simultaneously. For example, the sputtering yield of O atoms on SiC is proposed to be unity \citep{1994ApJ...431..321T}, which would lead to a significant destruction of their surface (because the abundance of O is high in nearby shells), enough to completely wipe the grains out. On the contrary, \cite{2003ApJ...598..785N} proposed a recycling scenario wherein the top 14\% of the surface is recycled multiple times while it stays in the SNe. This also explains the seemingly low concentrations of O atoms found in SiC grains\footnote{Other possible explanations proposed by \citealt{2003ApJ...598..785N} are the formation of O$_2$ or CO and diffusion through the grain}. Although SiC grains $\ga$ 0.1 $\micron$ can survive thermal sputtering in the remnant \citep{2016A&A...589A.132B}, they are prone to destruction due to non thermal sputtering by He$^+$ (Ar$^+$ and Ne$^+$ destroy SiO and other oxides formed in O rich zones in a similar manner) present in the ejecta. However, as shown in these studies, the sputtering yields of He, D, H and non thermal material on SiC is lower than that of O by at least an order of magnitude and SiC grains > 0.1$\micron$ can survive this destruction. Thus, we assume a 10\% surface destruction of the grains (\textit{i.e.,} loss of 10\% of the grain's surface area) leading to the loss of ions implanted in the top 10\% of the grain surface, which is also consistent with the 6-8\% surface erosion for $\micron$ sized C, Fe and Mg$_2$SiO$_4$ grains proposed by \cite{2007ApJ...666..955N}. 

\begin{figure}
\figurenum{2}
\label{fig:fig2}
\includegraphics[width=1.0\linewidth]{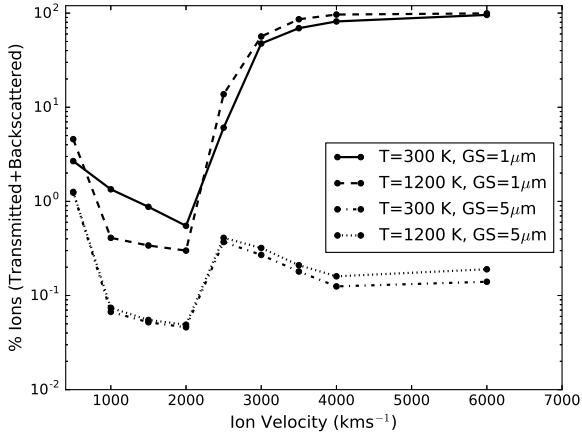}
\caption{Percentage of Cr ions transmitted and backscattered when incident with various velocities. GS refers to the grain size. Other ions show similar trends. Trends in implantation fraction for ion velocities $\le$ 2000 $\mathrm{km}\,\mathrm{s}^{-1}$ in a 1$\micron$ grain change when T > 800 K for Cr and Zn; however, larger grain shows no such variations in implantation fraction for all elements. Lower velocity region ($\le$ 1500 $\mathrm{km}\,\mathrm{s}^{-1}$) is dominated by backscattering and higher velocity region (> 2500 $\mathrm{km}\,\mathrm{s}^{-1}$) is dominated by transmission.}
\end{figure}

Transmission is the dominant phenomenon among BITS processes at ion velocities > 3000 $\mathrm{km}\,\mathrm{s}^{-1}$. At such velocities, the interaction time of the grain with ion is very less and the ion affects the spatial arrangement of target atoms only at highly oblique angles ($\ge$ 75$^{\circ}$). Based on our model, we find a high increment in the number of transmitted atoms (up to 40\% in certain cases) as compared to planar surfaces. Transmission is initially high at normal incident angles and is almost null beyond 70$^{\circ}$. As the incident velocity of ions is increased, transmission of ions starts at higher angles of incidence. Figure \ref{fig:fig2} summarizes the fraction of ions transmitted or backscattered against incident ion velocities, for different combinations of temperature and grain sizes. The trends shown are consistent for all the ions in question. We see a comparatively lower fraction ($\la$10\%) of ions backscattered and transmitted from the grain when ion velocities are < 3000 $\mathrm{km}\,\mathrm{s}^{-1}$. However, this shoots up to \textasciitilde50\% for velocities near 3000 $\mathrm{km}\,\mathrm{s}^{-1}$ and reaches nearly unity for velocities \textasciitilde6000 $\mathrm{km}\,\mathrm{s}^{-1}$. On the other hand, the 5$\micron$ grain hardly shows any transmission or backscattering and almost all the incident ions are implanted. Thus, we conclude there lies a certain range of ion velocities where implantation is dominant ($\sim$ 1000-2000 $\mathrm{km}\,\mathrm{s}^{-1}$). At velocities lower than this range, most ions are backscattered whereas for velocities higher than this range, most of them are transmitted.

We find that BITS processes are very sensitive to ion velocity and grain size, specially if the grain is smaller (\textasciitilde1$\micron$). For velocities $\la$ 2500 $\mathrm{km}\,\mathrm{s}^{-1}$, more than 90\% of the ions are implanted in a 1$\micron$ grain; for velocities between 2500-4000 $\mathrm{km}\,\mathrm{s}^{-1}$ and > 4000 $\mathrm{km}\,\mathrm{s}^{-1}$, 50-80\% and 90-97\% of ions are either transmitted or backscattered, respectively. We also see a drop of 7-87\% in implantation fraction at temperatures > 800 K for Cr and Zn ions, while implantation fraction for other three elements (Fe, Co, Ni) remain independent of temperature. For Zn, this can be attributed to its volatility (boiling point of Zn is 1180 K\footnote{\textit{Source:} Royal Society of Chemistry, www.rsc.org}) while for Cr, it can possibly be attributed to the formation of certain unstable complexes of Chromium and Carbon which could get evaporated at higher temperatures. These formations are highly favored if there is some Oxygen available as well such that CO-Cr complexes can be produced \citep{2013arXiv1308.4924S}. For Cr, this bias at high temperatures could also be simply due to Cr loosing its outermost electron in the \textit{4s$^1$} shell in the simulations. This would not be true for an SNe where highly ionized Cr would be present. The difference is then reflected in the final concentration estimated for these elements. For the 5 $\micron$ grain, implantation fraction remains constant for all temperatures $\le$ 2000 K \footnote{This observation was noted by running simulations at T = 1600 K and T = 2000 K whose results overlapped with the trends at T = 1200 K as shown in Figure \ref{fig:fig2}.}. 

\begin{figure}
\figurenum{3}
\center
\label{fig:fig3}
\includegraphics[width=1.0\linewidth]{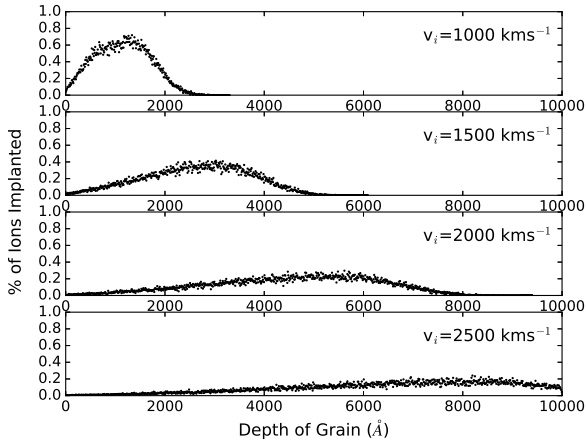}
\caption{Depth profiles obtained from SDTrimSP simulations for Cr ion relative velocities of 1000, 1500, 2000 and 2500 $\mathrm{km}\,\mathrm{s}^{-1}$ (subplots 1, 2, 3 and 4) incident on a $\micron$ sized SiC grain, plotted against \% of implantation. Simulation corresponding to 2000 $\mathrm{km}\,\mathrm{s}^{-1}$ shows maximal ion implantation in the core of the grain.}
\end{figure}

Simulations also predict changes in the spatial arrangement of atoms in the smaller grain's core due to ion implantation, specially when the population of ions implanted near the center of the grain is highest, as shown in Figure \ref{fig:fig3}, which has been reproduced from \cite{2017LPI....48.1490S}. For the larger grain, core implantation could not be achieved, even at velocities \textasciitilde6000 $\mathrm{km}\,\mathrm{s}^{-1}$. The effect on the core, although small, is significant and addresses the question of possibility of core contamination due to implantation in presolar grains. This is contrary to an often quoted assumption where core implantation is ruled out \citep{2003ApJ...594..312D,2008ApJ...689..622M}. Ions implanted in the core have a higher chance of survival and the signatures of smaller grains with sufficient core implantation can be preserved if they are embedded into bigger grains as subgrains. Thus, if an enhanced abundance of trace elements is obtained in X type grains with progenitor masses > 20 M\textsubscript{\(\odot\)}, a plausible conclusion of this excess abundance can be the inclusion of sub-grains within larger grains while they were still condensing. These subgrains may contain an enhanced abundance of trace elements condensed or implanted in their core, which could have survived after condensing in a bigger grain. As a matter of fact, recently analysis has found FeS and TiC subgrains in presolar grains \citep{2015PhDT........10G,2016ApJ...825...88H}. 

\subsection{Concentration Calculations}
\label{s:disc2}
We follow the work done by \cite{2003ApJ...594..312D} for calculating the concentrations of Cr, Fe, Ni, Co and Zn at different temperatures and velocities. The number of possible interactions that ions of a particular species in a column of uniform cross section (same as that of the grain) can have with the grain at a given time after explosion is described by: 

\begin{equation}
\label{eq:1}
\mathit{N_z^A} = \frac{N_A \sigma}{\mu_z} \int\limits_{m_0}^{M} \frac{X(m)  dm}{4 \pi r^2(m)} 
\end{equation}

where, $N_A$ is the Avogadro's Number, $\mu_z$ is the atomic weight of the element, $\sigma$ is the grain cross section, $r$ is the radial coordinate, $X(m)$ denotes the mass fraction of the isotope as a function of mass coordinate, $m_0$ denotes the mass coordinate at $r_0$ where $r_0$ is the grain condensation radius and $M$ is the total ejecta mass. Equation \ref{eq:1} does not consider the time expansion of ejecta. Hence, we introduce an additional term where the ratio of initial to final volume of the zones where implantation can take place is derived and used as a factor to decipher the possible number of interactions at a later time $t$. 

Early age supernovae have shock velocities of the order of a few thousand $\mathrm{km}\,\mathrm{s}^{-1}$ \citep{1991ApJ...375..652S,2017ApJ...837L...7B,2017ApJ...840..112S} and the differential zonal velocities can be taken to be a substantial fraction (\textasciitilde60\% and more) of the primary shock velocities because we consider the two outermost zones of the ejecta (He and H zones, respectively). For an estimate of final zonal widths, we consider a uniformly expanding ejecta with zones moving ahead with four possible differential zonal velocities ($\Delta$v) = 500, 1000, 2000 and 3000 $\mathrm{km}\,\mathrm{s}^{-1}$ respectively\footnote{The notion of a uniformly expanding ejecta will specially not hold if the shocked material is running into denser ambient material, like molecular clouds or is being shocked by the primary shock, cooled adiabatically and re$-$shocked by multiple reverse shocks (Sharda et. al. 2017, in preparation).}. We use the zonal nucleosynthesis yield model sets from S16 and apply the volume ratio factor to arrive at an estimate of number of possible interactions at times as late as a few hundred years (which marks the end of free expansion phase). To find the implantation concentration in ppm, we take its product with implantation fraction obtained using SDTrimSP at various velocities. 

For a typical explosion energy of $1.327\times10^{51}$ ergs, for a 15.2 M\textsubscript{\(\odot\)}, the ejecta mass as reported in S16 is 12.58 M\textsubscript{\(\odot\)} for which we obtain a radius of \textasciitilde4.54 pc and the time at which the free expansion phase ends as \textasciitilde431 years. Similarly, for 20.1 M\textsubscript{\(\odot\)} and 25.2 M\textsubscript{\(\odot\)} supernovae, the time marking the end of the first phase comes out as \textasciitilde466 and \textasciitilde458 years respectively. Although the progenitor masses differ by \textasciitilde25\% for the two heavier stars considered, their free expansion phase lifetime is similar, perhaps because the stellar winds reduce the heavier star (prior to explosion) to the size same as that of the lighter star \citep{2009ARA&A..47...63S}. We also observe that the ejecta mass thrown out by both the stars is same (\textasciitilde15 M\textsubscript{\(\odot\)}), which reinforces the argument made above. 

We describe the calculations for $^{52}$Cr here, at various velocities and temperatures for 1$\micron$ and 5$\micron$ SiC grains for 15.2 M\textsubscript{\(\odot\)}, 20.1 M\textsubscript{\(\odot\)} and 25.2 M\textsubscript{\(\odot\)} stars, respectively. From the data for 15.2 M\textsubscript{\(\odot\)}, it is observed that for a condensation radius of $1.295\times10^6$ km (mid-point of He zone), $X(m)$  is a constant for $^{52}$Cr. The ppm concentration (by weight) is given by:

\begin{equation}
\label{eq:2}
\mathit{I_c(ppm) = 10^6 I_i  N^A_z \mu_z \frac{d^3_M - d^3_{m_0}}{{(t \Delta v)}^3\frac{4 \pi}{3}r^3_{SiC}\rho_{SiC}}} 
\end{equation}
        
where, $I_i$ is the implantation fraction of ions and is a function of ion velocity (relative to grain), grain cross section, grain density and temperature. From a numerical integration of equation \ref{eq:1} with $m_0$ = 3.95 and $M$ = 12.58, for a 1$\micron$ grain, we get $N^A_z$ = 1.67$\times10^9$.  Thus, for a differential zonal velocity ($\Delta$v) of 2000 $\mathrm{km}\,\mathrm{s}^{-1}$ between zones 499 (mid-point of He zone) and 950 (last zone) in S16 zonal yield sets, at a time $t\,$(days) after explosion, the ppm concentration (by weight) is given by: 

\begin{equation}
\mathit{I_c  (ppm) = \frac{\beta I_i}{t^3 (days)}} 
\end{equation}

where, $\beta$ is a constant. Assuming grain condensation is fully achieved by $t$=1700 days \citep{2017ApJ...843...57D} and the free expansion phase ends around 430 years, we get a ppm concentration of \textasciitilde0.62 for $^{52}$Cr implanted in a 1$\micron$ grain when its velocity is 1000 $\mathrm{km}\,\mathrm{s}^{-1}$. 

So far, the mass fraction has been assumed to be constant at $t$ > 200 seconds. However, two bipolar mechanisms can alter this mass fraction: the addition of mass into zones which are farther out as the ejecta sweeps up material from ambient ISM and the addition of mass from zones interior to the condensation zone because of RT instabilities (mixing) which are caused when a lighter fluid tends to push over a heavier fluid (discussed in detail in section \ref{s:disc3}). The amount of mixing remains a largely unsettled question, however, to begin with, we consider a 1\% mixing between He/C and He/N zones (as estimated by NanoSIMS analysis of presolar grains by \citealp{2008ApJ...689..622M}), \textit{i.e.}, we add 1\% of the zonal yields from He/C zone to the He/N zone while calculating implanted concentrations. The mass fraction of $^{52}$Cr can increase to around double the value of $X(m)$ taken in this calculation (\textasciitilde83\% increase in the case of $^{52}$Cr implanted into SiC condensed in a 15.2M\textsubscript{\(\odot\)} SNe). The free expansion phase ends when the SN ejecta has swept up a mass $\sim\,12.6$ M\textsubscript{\(\odot\)}. It can be assumed without the loss of generality that the swept up mass which affects the mass fraction of ions in question is at least 60\% of the total mass of the expanding supernova (for e.g., the swept up mass of Tycho's supernova remnant (SNR) present in the outermost regions is \textasciitilde53\% of the total mass of the SNR, as derived by \citealt{1983ApJ...266..287S}). Anyway, this factor is suppressed when mixing from inner regions is taken into account, as we explain in section \ref{s:disc3}. It can also be safely assumed that there was no production of Cr before the explosion and hence the swept up mass does not contain any significant quantity of the isotope\footnote{The same assumption is valid for other four ions because the contribution to their yields by the winds before explosion is negligible as compared to their abundances produced after explosion (S16).}. Thus, the mass fraction decreases by a factor
\begin{equation}
\label{eq:4}
\Delta X(m) = \frac{12.58-3.95}{(12.58-3.95) + (0.6\times12.58)} = 53.35\% 
\end{equation}

Overall, we get a \textasciitilde30\% rise in the mass fraction of Cr, which gives a corrected value of \textasciitilde0.80 ppm. In an attempt to account for grain destruction in later years in the ejecta and the ISM, we assume the loss of top 10\% of the grain surface, as noted in section \ref{s:disc1}. Utilizing the depth profiles obtained through spherical grain approximation for each ion implanted in the grain, we find that for this particular case, \textasciitilde59\% of the ions implanted are able to penetrate the grain to more than 10\% of the grain's depth. The surface erosion corrected ppm concentration thus obtained is \textasciitilde0.79 ppm. This concentration still lacks mixing from innermost zones and additional contributions from radioactive nuclei, which we correct for in section \ref{s:disc3}. Interestingly, for the larger grain, most of the ions implanted are in layers deeper than the $\sim$10\% erosion barrier threshold we work with and are hence preserved during surface erosion.

\subsection{Comparison With Laboratory Measurements}
\label{s:disc3}
The calculated concentrations of transition ions implanted in the grains vary over 1-3 orders of magnitude when ion velocities and differential zonal velocities are varied between 1000-6000 and 500-4000 $\mathrm{km}\,\mathrm{s}^{-1}$ respectively. However, not all the calculated values would correspond to physically plausible initial conditions, so we maintain a check on those sets which yield erroneously high concentrations. 

Laboratory measurements of these concentrations include contribution from both processes: condensation and implantation\footnote{While laboratory measurements from other instruments may contain errors due to sample contamination \citep{2006ApSS..252.7117H,2008LPI....39.2135K} those from NanoSIMS are able to avoid it and have been shown to be quite accurate \citep{2006M&PSA..41.5248M,2006LPI....37.1355H}}. The estimates of concentrations via condensation and implantation can be matched with those measured in the grains. X type SiC grains from the Murchison meteorite analyzed in \citet{2008ApJ...689..622M} have a mean size of \textasciitilde2.5$\micron$ with 90\% sizes lying in the range 1.8-3.7$\micron$ \citep{1994ApJ...430..870H,ZINNER20074786} whereas those analyzed by \cite{2000MPS...35.1157H} have grain sizes in the range 0.5-1.5 $\micron$. To compare the concentrations measured by \cite{2008ApJ...689..622M} with our calculations, we require the implantation fraction for each incident element. We approximate it by taking a geometric mean of the resultant implantation fractions for the two grain sizes we study. 

Laboratory based ppm concentrations of Fe and Ni range from a few tens to a few thousands for SiC X grains found in Murchison meteorite while those of Co are mostly a few tens of ppm. The 15 M\textsubscript{\(\odot\)} supernova model is believed to better explain these abundances than the 25 M\textsubscript{\(\odot\)} model. as stated by \cite{2008ApJ...689..622M}, however the models these authors used were taken from \cite{2002ApJ...576..323R} which have been improved upon in S16. Concentrations of Fe in smaller SiC X grains studied by \cite{2000MPS...35.1157H} lie in the range \textasciitilde100-1000 ppm. Our calculated values for Fe implanted in the grains is 0-2 orders of magnitude lesser (on average\footnote{This average is the mean of concentrations measured by \cite{2008ApJ...689..622M}.}) for ion velocities in 1000-3000 $\mathrm{km}\,\mathrm{s}^{-1}$ and differential zonal velocities between 1000-3000 $\mathrm{km}\,\mathrm{s}^{-1}$. We are not aware of any measured concentrations for Zn found in presolar SiC X grains. Though Cr concentrations have been measured to be $\sim$1 ppm in SiC X grains (measured by \cite{2008LPI....39.2135K}, as reported in \citealt{2009IJMSp.288...36L}), we cannot utilize them since they do not belong to the same grains as analyzed by \cite{2008ApJ...689..622M}. Hence, we leave these elements out in the discussion for comparison but we propagate the effects of various mixing criteria we have considered in them.

Contrary to finding excess Ni in type X presolar SiC grains (<Fe/Ni> = 0.78 (0.14, 3.34) for the SiC X grains analyzed by \cite{2008ApJ...689..622M}), we find Fe/Ni >1. Ion implantation as a probable cause for this excess was ruled out by \cite{2008ApJ...689..622M} because these excesses were distributed all across the grain instead of being localized in the outermost regions and core implantation was not taken into account. This motivates us to consider mixing from the innermost regions (Si/S, Ni zones) as well, which are rich in Ni and contributions from these regions can possibly explain the observed excess of Ni. 

It has been shown that concentrations of certain isotopes of Si, Ti and Ca obtained in laboratory measurements of Carbide (Graphite and SiC) grains can only be explained if there is intense mixing between inner Si/S zones and outer He zones \citep{1999ApJ...510..325T,2002ApJ...576L..69H} while the same has been postulated for explaining excess Ni obtained in these measurements \citep{2012ApJ...758...59S}. This can happen through Si rich jets originating from Si zone in the interior and cutting across O rich zones, throwing material from the inner regions all the way out to He and H zones. The presence of Si rich jets owing to an asymmetric explosion \citep{1999ApJ...524L.107K} has been often reported (\citealp{2004ApJ...615L.117H,2017MNRAS.468.1226G} and references therein) which support the theory of mixing from innermost zones to sites of carbide grain condensation in outer zones \citep{2006ApJ...647L..37L}. These jets also cause $\alpha$ rich freeze-out behind the energetic shock which is essential for the production of transition elements in question \citep{1997ApJ...486.1026N,2000ApJS..127..141N} and their presence would also constrain the mixing from intermediate O rich zones to lower values which is necessary in order to limit the amount of Oxygen available in the grain surroundings so that C/O >1 is preserved, O is held up in CO and oxide formation can be suppressed. A homogeneous mixing from all zones is thus not preferred for explaining observed elemental abundances in presolar grains \citep{2004ASPC..309..265H}. 

3D simulations of mixing in the ejecta performed by \cite{2010ApJ...714.1371H} predict the formation of 'bullets' (clumps) of Z >8 elements (called Ni rich bullets in their paper), some of which are fast enough to overtake the O rich bullets and reach the outer He and H zones within the first 10000 seconds of explosion. The same has been observed in SN1987A \citep{1989ApJ...341L..63A}. This decay can take place in the innermost shell which is moving the slowest and as the ejecta cools adiabatically, $\gamma$-rays from this decay cause local heating which sends out a pressure wave towards the outside, thus giving rise to conditions necessary for RT mixing. 3D simulations of CCSN predict mixing to cease by \textasciitilde10$^5$ seconds for a 15 M\textsubscript{\(\odot\)} star \citep{2010ApJ...723..353J}.

Keeping these studies in mind, we consider a 1\% contribution (through mixing) of Ni from Si/S zone and add contributions from those radioactive ions which may travel to outer zones along with Ni and Si rich 'bullets'  \footnote{This is unlike the algorithm used by \citet{2007ApJ...666.1048Y} to explain observed isotopic ratios (of same elements) in SiC X grains by manipulating mixtures from different zones according to each individual grain.}. Specifically, we trace $^{52}$Ni, $^{52}$Fe and $^{52}$Mn for $^{52}$Cr, 2.) $^{56}$Co, $^{56}$Ni for $^{56}$Fe and 3.) $^{59}$Ni for $^{59}$Co\footnote{See National Nuclear Data Centre, Brookhaven National Laboratory's website for details on decay times and decay cascades: https://www.nndc.bnl.gov/ensdf/}. Amongst these, we neglect $^{52}$Fe, $^{52}$Mn and $^{52}$Ni because their zonal contributions are negligible as compared to that of their end products in the zones of interest. The half life of $^{59}$Ni is > 10$^4$ years which implies that only 0.4\% of $^{59}$Ni has decayed into $^{59}$Co by the end of free expansion phase \citep{RUHM1994227}. Thus, we use a 1\% fraction of the decayed 0.4\% $^{59}$Ni from the innermost Ni zone and maintain the same fraction for all other elemental yields taken into account from this zone so that zonal mixing is same for all elements and calculations are unbiased in every zone (\textit{i.e.}, elemental fractionation is not favored by ion implantation).

Consequently, we find that the concentrations of Ni in a 1$\micron$ grain can increase by 19$\times$, 14$\times$ and 8$\times$ for 15.2, 20.1 and 25.2 M\textsubscript{\(\odot\)} models respectively, with respect to the concentrations calculated before this mixing is taken into account. Similarly, the concentrations of Co increase by \textasciitilde2.5$\times$ for all the three models whereas the concentrations of Cr increase by 5.2$\times$, 2.3$\times$ and 1.8$\times$ respectively. Zn remains unaffected by mixing from the interior. This additional concentration for Ni is solely from mixing whereas the increments in concentrations of Cr and Co come from mixing as well as radioactivity corrections. Contributions to $^{56}$Fe from $^{56}$Co do not lead to significant increments. The production of $^{56}$Ni has been the focus of all supernovae nucleosynthesis models and while it is a key factor to uncovering the mysteries of supernovae explosions (\citealt{1989A&A...210L...5H,1990ApJ...349..222T,2002RvMP...74.1015W,2006NuPhA.777..424N,2007PhR...442..269W,2015ApJ...807..110J}, S16, \citealt{2017arXiv170404780S}), its contribution to the ppm concentrations of $^{56}$Fe make Fe/Ni high ( $\ge\,$1) whereas Fe/Ni <1 has been measured in \textasciitilde73\% of all the SiC X grains analyzed by \cite{2008ApJ...689..622M}. Thus, we consider two scenarios: one where we refrain from adding significant contributions from $^{56}$Ni to our calculated concentrations of $^{56}$Fe and the other with taking it into account. The former is motivated by 3D simulations  of supernovae explosions which predict that most of the mass of $^{56}$Ni resides in two big clumps moving in opposite directions \citep{2013A&A...552A.126W}. This is also in concurrence with the observed structure of SN1987A 23 years after the explosion by \cite{2016ApJ...833..147L} where the authors quote that although the 3D simulation by \cite{2013A&A...552A.126W} models the SNe environment a few hundred seconds after the explosion, the overall structure and spatial distribution of $^{56}$Ni should hardly change in subsequent times. Similarly, presence of bipolar $^{56}$Ni jets have been detected in SNR 2013ej which is a strong evidence for inhomogeneous and clumpy distribution of this isotope \citep{2017MNRAS.472.5004U}. We thus assume that the site of SiC condensation is away from these high velocity clumps of $^{56}$Ni. In any case, most of these high velocity (4000-6500 $\mathrm{km}\,\mathrm{s}^{-1}$) $^{56}$Ni ions would simply traverse the grain without significant implantation. For the latter scenario where we assume the formation site and consequent movement of SiC grains near $^{56}$Ni clumps, we assume a 0.001\% mixing of this isotope for the production of $^{56}$Fe, in line with the work of \cite{1998MNRAS.299..150F} where this amount of mixing of $^{56}$Ni from innermost Ni zone to outer He zones has been used to reproduce the observed He I line in SN1995V.

\begin{deluxetable}{|c c c c c c|}
\tiny
\tabletypesize{\scriptsize}
\tablenum{1}
\tablecolumns{6}
\tablecaption{Relative elemental abundances (with respect to Fe) calculated in this work with contributions from both implantation and condensation compared with experimental results. Values in column 3 are means of the relative abundances measured in different grains, with the lowest and highest relative abundances shown in parenthesis. The table is divided into two parts which show relative abundances when 2\% and 4\% mixing is considered from Si/S zone respectively. For each of these parts, values in first row denote the maximum and minimum ratios obtained when yields from $^{56}$Ni are not added to $^{56}$Fe; whereas those in second row denote relative abundances when this yield is taken into account.}
\tablehead{\colhead{Elements} & \colhead{Solar\tablenotemark{a}} & \colhead{Experimental} & \colhead{15M\textsubscript{\(\odot\)}} & \colhead{20.1M\textsubscript{\(\odot\)}} & \colhead{25.2M\textsubscript{\(\odot\)}}}
\startdata
\multicolumn{6}{|c|}{\textit{Si/S Zone Mixing = 2\%}}\\
Fe/Cr&66.0&....&14-26&44-76& 41-70\\
&&&18-34&49-86&45-77\\
Fe/Co&362.9&31 (3-80)\tablenotemark{b}& 32-84&76-200& 58,153\\
&&&41-112&89-141&70-187\\
Fe/Ni&17.8&0.8 (0.1- 3.3)\tablenotemark{b}&0.90-1.03&1.43-1.61&1.6-1.8\\
&&&1.24-1.34&1.51-1.68&1.74-1.97\\
Fe/Zn&690.9&....&790-834& 2098-2215&1282-1353\\
&&&1017-1073&2342-2517&1425-1657\\
\hline
\hline
\multicolumn{6}{|c|}{\textit{Si/S Zone Mixing = 4\%}}\\
Fe/Cr&66.0&....&15-29&46-83& 43-74\\
&&&21-39&53-111&49-96\\
Fe/Co&362.9&31 (3-80)\tablenotemark{b}& 35-92&81-227&62-161\\
&&&47-129&94-231&69-185\\
Fe/Ni&17.8&0.8 (0.1-3.3)\tablenotemark{b}&0.40-0.50&0.77-0.87&0.91-1.03\\
&&&0.63-0.69&0.84-0.99&1.04-1.13\\
Fe/Zn&690.9&....&869-917&2218-2435&1371-1467\\
&&&1170-1234&2584-2738&1574-1790\\
\hline
\enddata
\tablenotetext{a}{\citet{2003ApJ...591.1220L}}
\tablenotetext{b}{\citet{2008ApJ...689..622M}}
\label{tab:tab1}
\end{deluxetable}

However, despite the above additions from inner zones and radioactivity corrections, we fail to cover the whole range of measured Fe/Ni and Fe/Co ratios. On experimenting further with different sets of mixing, we find that a 2\% mixing from Si/S zone (instead of the 1\% considered so far, keeping other mixing contributions constant) simultaneously generates the desired abundances of Ni and Co (relative to Fe) to a certain extent. To get to the lowest Fe/Ni ratios reported in \cite{2008ApJ...689..622M}, we find that a higher contribution ($\ge$ 4\%) is required from the Si/S zone because in this zone, while $^{58}$Ni is still in excess, $^{56}$Ni has highly depleted from its value in the innermost Ni zone. Thus, we land at our final calculated values of condensed as well as implanted concentrations for the species of interest by assuming 1.) 0.004\% and 0.001\% contribution for $^{59}$Ni and $^{56}$Ni (respectively), 2\% and 1\% mixing from Ni, Si/S and He/C zones and 2.) 0.004\% and 0.001\% contribution for $^{59}$Ni and $^{56}$Ni (respectively), 4\% and 1\% mixing from Ni, Si/S and He/C zones, for a SiC X grain formed in the He zone. We summarize the relative abundances obtained by modeling implantation$+$condensation in Table \ref{tab:tab1} for the two scenarios put forth. The difference between the two scenarios can be attributed to SiC grains condensing and evolving near or far from Si rich ejecta present in the outermost layers. However, since the condensation of SiC is highly favorable if Si rich clumps (ejected outwards from inner Si rich regions; see Figure \ref{fig:fig1} where Si/C >> 1) are present, scenario 2 seems more probable. Thus, mixing from inner zones can also provide an explanation for the high isotopic abundances of Si in SiC X grains. If the percentage of mixing is increased by 2$\times$ in either the Ni or the He zones, it leads to an overproduction of elemental abundances by implantation itself, without leaving room for condensation. On the other hand, the concentrations decrease by 40\% when mixing is taken to be 0.5\% in He/C zones, which can be considered a lower threshold since a value less than this will not produce enough abundance through implantation in the grains. 

During the first few hundred years after explosion, the differential zonal velocity between He and H zones is of the order of a few thousand $\mathrm{km}\,\mathrm{s}^{-1}$. Thus, the majority of implantation during the free expansion phase should come when $\Delta$v > 1000 $\mathrm{km}\,\mathrm{s}^{-1}$. There is a possibility of higher differential zonal velocities than considered in our work but they would not exist for a long time (as compared to the timeline of few hundred years we use). Also, their contribution to the implanted concentrations would be scarce as compared to the ones considered in our calculations. We also consider a case with $\Delta$v = 500 $\mathrm{km}\,\mathrm{s}^{-1}$, however, the concentrations we calculate are higher than measured in more than 80\% of the grains which leads us to reject this set in most comparisons. With this view, we find a probable range of fraction of abundances implanted in the SiC X grains for which ppm concentrations of Fe, Co and Ni have been measured by \cite{2000MPS...35.1157H} and \cite{2008ApJ...689..622M}. 

In Tables \ref{tab:tab21} and \ref{tab:tab22}, we present the maximum fraction of these elements which can come from implantation when zonal mixing from Si/S zone is 2\% and 4\% respectively, by taking geometrical mean of the implanted concentrations we find for the two grain sizes (since the average size reported in analysis of SiC X grains is \textasciitilde2.5 $\micron$). We reject certain sets which show an implantation fraction > 1 for all combinations of parameters and we call the case 'NP' (Not Possible). For grains which show lesser concentrations, we assume they have been ejected into the ISM earlier than others. By 'early' ejection, we imply that the grain gets out of the reach of high velocity ions in the shocked ISM earlier than its expected ejection time into ambient ISM. This 'early' ejection scenario is possible if the SiC grain condenses near Si rich ejecta clumps moving outwards at high velocities because such clumps can cross the forward shock and move ahead of it; essentially imitating an early ejection. As a matter of fact, Si rich clumps have been observed to be moving ahead of the forward shock in the Vela CCSN \citep{2017A&A...604L...5G}. However, this mechanism requires stark density contrasts between the clump and its surroundings. Also, the time it takes for the clump to overtake the forward shock is not known with surety (see also \citealt{1988LNP...316.....K,1995Natur.373..587A}). Another way the early ejection could be achieved is through the presence of a huge shock wave which accelerates the dust and not the gas around it. However, the origin of such a shock wave remains unclear. Thus for such grains we only consider implantation at the earliest epochs when differential zonal velocities were the highest. To explain higher concentrations, we subsequently include contributions from lower differential zonal velocity sets while assuming that these grains spent a longer time in the SNe. 

As is seen from Tables \ref{tab:tab21} and \ref{tab:tab22}, implantation fraction predicted for these elements covers the whole range from 0-1 depending on the physical conditions present in the SNe. If we were to believe that implantation should not contribute more than 60\% (the highest predicted implantation fraction for heavy elements, \citealt{2004ApJ...607..611V}) to the total concentration of transition elements found in the grain, it would imply that most of the ions are implanted in the free expansion phase when $\Delta$v is still a few thousand $\mathrm{km}\,\mathrm{s}^{-1}$. Since the shock velocities are of the order of a few thousand $\mathrm{km}\,\mathrm{s}^{-1}$ in this phase, the argument made above supports using $\Delta$v > 1000 $\mathrm{km}\,\mathrm{s}^{-1}$ as most probable differential zonal velocities because at one end of it, we have nothing but the shock velocity (since we deal with the outermost layers of the ejecta). Moreover, we observe that there is almost always a steep decline in implantation fraction as one moves from a zonal velocity of 1000 to 2000 $\mathrm{km}\,\mathrm{s}^{-1}$. Since this decline does not seem continuous, it becomes straightforward to demarcate a maximum possible implantation fraction for the elements we study in SiC X grains. 

Thus, if we only take into account $\Delta$v > 1000 $\mathrm{km}\,\mathrm{s}^{-1}$, the implantation fraction we get is $\la$ 0.25 for grains condensed in 15 M\textsubscript{\(\odot\)} model, where measured concentrations of Fe and Ni are $\ga$ 300 ppm. For lower concentrations of Fe and Ni, this fraction could reach as high as \textasciitilde60\% while for measured concentrations $\ga$ 1000, this fraction drops below 0.1, implying condensation is the dominant process among the two unless the grain did not spend much time in the SNe, as postulated earlier. The model also predicts that if these SiC grains were synthesized in heavier stars ($\ge$ 20 M\textsubscript{\(\odot\)}), they would have spent a lesser effective time in the SNe or the zonal velocities would have been higher in the free expansion phase. Although we assume fixed zonal velocities throughout this phase, one can achieve more accuracy in concentrations by taking appropriate fractions of each zonal velocity yield. In most cases, for identical conditions, implantation fraction of Ni is thought to be more than Fe which makes sense because Ni is more volatile than Fe so condensation fraction for Fe should be higher if same amounts of Fe and Ni condense with the grain.

Concentration of other two elements - Cr and Zn, is relatively lower - of the order of 1 and 0.1 ppm for a 15 M\textsubscript{\(\odot\)} model, 10 and 0.3 ppm  for a 20 M\textsubscript{\(\odot\)} model and 40 and 3 ppm for a 25 M\textsubscript{\(\odot\)} model. A substantial amount of Zn found in SiC X grains should come from implantation and not condensation because of its volatile nature and hence making it difficult to co-condense with the grain. On the other hand, Cr is refractory so a substantial amount of it can also come from condensation. These elements would be discussed in detail in a future work wherein measured concentrations from other X type grains will be available. We also leave calculations for Ti and V (whose ppm concentrations have been measured in presolar grains by \citealt{2001LPI....32.2192K,2002LPI....33.2056K}) for a future work.  

\begin{deluxetable*}{|c| c c c c c | c| c c c c c|}
\tiny
\label{tab:tab21}
\tabletypesize{\scriptsize}
\tablenum{2}
\tablecolumns{12}
\tablecaption{Measured ppm concentrations from \cite{2008ApJ...689..622M} and proposed maximum percentage of ion implantation, unless the maximum is 100\%, in which case the minimum percentage is shown. Mixing set used is: 0.001\% and 0.004\% (for $^{56}$Ni and $^{59}$Co) from Ni zone, 2\% from Si/S zone and 1\% from He/C zone. $\Delta$v denotes differential zonal velocities (in $\mathrm{km}\,\mathrm{s}^{-1}$) between He and H zones. For the three supernovae models for $^{56}$Fe, values in first row denote the maximum implantation fraction obtained when yields from $^{56}$Ni are not added to $^{56}$Fe; whereas those in second row denote the maximum fraction of ions implanted when this yield is taken into account. When implantation fraction > 1 for all parameter sets in a model, the model is marked 'NP' (Not Possible). For largest measured concentrations, the sample size is 100\% hence no appropriate groups of measured concentrations could be made. Grain size has been assumed to be \textasciitilde2.4$\micron$. As long as T $\le$ 2000 K, implantation in Fe, Co and Ni remain independent of temperature.}
\tablehead{\colhead{Element} & \colhead{Conc$_{measured}$} & \colhead{$\Delta$v} & \colhead{\% I$_{max}$ 15 M\textsubscript{\(\odot\)}} & \colhead{\% I$_{max}$ 20 M\textsubscript{\(\odot\)}} & \colhead{\% I$_{max}$ 25 M\textsubscript{\(\odot\)}} \vline& \colhead{Element} & \colhead{Conc$_{measured}$} & \colhead{$\Delta$v} & \colhead{\% I$_{max}$ 15 M\textsubscript{\(\odot\)}} & \colhead{\% I$_{max}$ 20 M\textsubscript{\(\odot\)}} & \colhead{\% I$_{max}$ 25 M\textsubscript{\(\odot\)}} }
\startdata
Fe&$\le$ 50&3000&75&NP&NP&Co&$\le$ 20&1000&58&$\ge$ 81&NP\\
&&&96&NP&NP&&&2000&7&24&$\ge$ 43\\
&50-150&2000&62&NP&NP&&&3000&2&7&30\\
&&&78&NP&NP&&25-70&500&$\ge$56&NP&NP\\
&50-150&3000&14&$\ge$ 78&NP&&&1000&17&55&NP\\
&&&32&$\ge$85&NP&&&2000&2&7&29\\
&150-310&2000&40&NP&NP&&&3000&0.6&2&9\\
&&&52&NP&NP&&200-220&500&42&$\ge$ 59&NP\\
&&3000&12&93&NP&&&1000&0.5&17&73\\
&&&15&$\ge$59&NP&&&2000&0.6&2&9\\
&310-640&1000&$\ge$ 64&NP&NP&&&3000&0.1&0.6&3\\
&&&NP&NP&NP&Ni&$\le$ 100&2000& $\ge$ 52&NP&NP\\
&&2000&20&$\ge$ 62&NP&&&3000&37&$\ge$ 77&NP\\
&&&25&NP&NP&&100-250&2000&50&NP&NP\\
&&3000&6&45&NP&&&3000&15&73&$\ge$59\\
&&&7&70&NP&&250-600&1000&70&NP&NP\\
&900-1100&1000&91&NP&NP&&&2000&21&$\ge$ 43&NP\\
&&&$\ge$ 48&NP&$\ge$50&&&3000&6&31&89\\
&&2000&11&89&NP&&650-900&1000&$\ge$ 47&NP&NP\\
&&&14&$\ge$56&NP&&&2000&14&68&$\ge$ 84\\
&&3000&3&26&85&&&3000&4&20&60\\
&&&4&41&NP&&950-1200&1000&83&NP&NP\\
&1100-1800&1000&56&NP&NP&&&2000&10&81&$\ge$ 63\\
&&&72&NP&NP&&&3000&3&24&45\\
&&2000&7&54&$\ge$71&&1300-1700&1000&59&NP&NP\\
&&&9&85&NP&&&2000&7&57&44\\
&&3000&2&16&52&&&3000&2&17&32\\
&&&3&25&75&&2000-2500&1000&40&NP&NP\\
&2000-3000&1000&33&NP&NP&&&2000&5&39&71\\
&&&43&NP&NP&&&3000&1&12&21\\
&&2000&4&32&$\ge$43&&\textasciitilde3000&1000&33&NP&NP\\
&&&5&51&62&&&2000&4&32&60\\
&&3000&1&10&31&&&3000&1&10&18\\
&&&2&15&45&&\textasciitilde3300&1000&30&NP&NP\\
&\textasciitilde3500&1000&29&NP&NP&&&2000&4&29&54\\
&&&37&NP&NP&&&3000&1&9&16\\
&&2000&5&28&90&&\textasciitilde4500&500&$\ge$ 74&NP&NP\\
&&&5&44&53&&&1000&22&$\ge$ 70&NP\\
&&3000&1&8&27&&&2000&3&22&40\\
&&&1&13&39&&&3000&0.8&6&12\\
&\textasciitilde4500&500&$\ge$ 73&NP&NP&&\textasciitilde5400&500&$\ge$ 62&NP&NP\\
&&&$\ge$ 93&NP&NP&&&1000&18&$\ge$ 59&NP\\
&&1000&22&$\ge$ 70&NP&&&2000&2&18&33\\
&&&29&NP&NP&&&3000&0.6&5&10\\
&&2000&3&22&70&&&&&&\\
&&&4&34&41&&&&&&\\
&&3000&0.8&6&21&&&&&&\\
&&&1&10&30&&&&&&\\
\hline
\enddata
\end{deluxetable*}

\begin{deluxetable}{|c| c c c c c|}
\tiny
\label{tab:tab22}
\tabletypesize{\scriptsize}
\tablenum{3}
\tablecolumns{6}
\tablecaption{Same as Table \ref{tab:tab21}, but zonal mixing from Si/S zone is increased to 4\%. The only significant impact is caused on implantation fractions of Ni (those of Fe increase by \textasciitilde8\% while they do not change for Cr, Co and Zn). Some sets which were included in Table 2.1 have been removed since all models were termed NP.}
\tablehead{\colhead{Element} & \colhead{Conc$_{measured}$} & \colhead{$\Delta$v} & \colhead{\% I$_{max}$ 15 M\textsubscript{\(\odot\)}} & \colhead{\% I$_{max}$ 20 M\textsubscript{\(\odot\)}} & \colhead{\% I$_{max}$ 25 M\textsubscript{\(\odot\)}} \vline}
\startdata
Ni&$\le$ 100&3000& $\ge$ 35&NP&NP\\
&100-250&2000&$\ge$47&NP&NP\\
&&3000&33&96&NP\\
&250-600&2000&47&86&NP\\
&&3000&14&55&NP\\
&650-900&2000&31&83&$\ge$ 84\\
&&3000&9&36&$\ge$\\
&950-1200&1000&$\ge$79&NP&NP\\
&&2000&23&$\ge$53&NP\\
&&3000&7&48&93\\
&1300-1700&1000&$\ge$56&NP&NP\\
&&2000&16&78&$\ge$69\\
&&3000&5&31&50\\
&2000-2500&1000&90&NP&NP\\
&&2000&11&62&$\ge$47\\
&&3000&3&29&34\\
&\textasciitilde3000&1000&75&NP&NP\\
&&2000&9&74&93\\
&&3000&3&27&28\\
&\textasciitilde3300&1000&68&NP&NP\\
&&2000&9&67&85\\
&&3000&3&39&25\\
&\textasciitilde4500&1000&50&NP&NP\\
&&2000&6&47&62\\
&&3000&2&21&19\\
&\textasciitilde5400&1000&42&NP&NP\\
&&2000&5&51&52\\
&&3000&2&47&16\\
\hline
\enddata
\end{deluxetable}

A similar analysis (as we perform in this work) was recently carried out by \cite{KODOLANYI2017} where the authors measured concentrations of Fe and Ni isotopes in SiC X grains obtained from KJD \citep{1994GeCoA..58..459A} and Mur2012B \citep{2014LPI....45.1031H} grain separates of the Murchison meteorite. Although they find Fe/Ni concentrations in different SiC X grains to vary over two orders of magnitude (from 0.36 to 37.6), their measurements are doubtful due to possible contamination from multiple phases while sample preparation because the grains they analyzed were smaller and their diameter was comparable to the beam diameter of the desorption laser used. Keeping this in mind, the authors further discussed the concentrations of only three particular grains which they expected to be least affected by contamination (see section 4.2 in their paper). The first of these belonged to the KJD mount whereas the rest two belonged to the Mur2012B mount. Furthermore, the sizes of these three grains are between $\sim$0.5-1.0 $\micron$ which is similar to the ones we have simulated. For these three grains, the authors attempted to establish links with CCSN nucleosynthesis models through a variety of mixtures of elements from different zones. The nucleosynthesis models they used were from \cite{2002ApJ...576..323R} and \cite{2015ApJ...808L..43P} where the latter model included ingestion of abundances from the outermost H zone. However, they were not able to reproduce the desired abundances of all the isotopes while maintaining C/O >1 from either of these models, for two of the three grains. On comparing the Fe/Ni ratios they find for the three grains (0.36, 1.26 and 1.24 respectively), we immediately see that we are able to reproduce the measured Fe/Ni ratios in them using the same nucleosynthesis model (15 M\textsubscript{\(\odot\)} and mixing criteria we have considered for grains analyzed in \cite{2008ApJ...689..622M}; see Table \ref{tab:tab1}). This is encouraging and highly expected because the grains belonged to the same meteorite. Thus, it is logical to argue that grains embedded in different parts of the same meteorite originated from the same CCSN.

\section{Summary}
\label{s:summ}
We have developed a theoretical model to estimate the fraction of transition elements condensed and implanted in SiC X grains and compared it with their concentrations obtained from SiC X grains found in Murchison meteorite. For this calculation, we analyzed ion-grain interactions at various sets of relative velocities, zonal velocities and temperature using an ion target simulator SDTrimSP. We use nucleosynthesis zonal yield sets generated by S16 for 15, 20 and 25 M\textsubscript{\(\odot\)} stellar models. We also take into account the time expansion of ejecta through the free expansion phase and associate appropriate radioactive corrections to our analytical calculations. This model is fairly versatile due to its two degrees of freedom (namely, ion velocity and differential zonal velocity) and can be applied to calculate implanted concentrations of all other elements (and respective isotopes) in the X grains condensed in SNe for all types of progenitor masses. Our main conclusions are as follows:

\begin{enumerate}
\item Backscattering is only effective for lower ion velocities (v < 1000 $\mathrm{km}\,\mathrm{s}^{-1}$) and highly oblique angles. Only $\la$ 6\% of the ions incident are backscattered for the range of velocities we consider in this work. Using our geometric model described in Appendix \ref{s:append}, we find transmission to dominate implantation for ion velocities > 4000 $\mathrm{km}\,\mathrm{s}^{-1}$ for SiC grains of size 1$\micron$. Maximum transmission is $\le$ 10\% for velocities $\le$ 2500 $\mathrm{km}\,\mathrm{s}^{-1}$, after which it shoots up to > 50\%, reaching almost unity for v > 4000 $\mathrm{km}\,\mathrm{s}^{-1}$. We also confirm that sputtering of Si and C atoms by transition elements is largely ineffective.
\item While the implantation of transition elements remain fairly independent of temperature (provided T $\le$ 2000 K) for Fe, Co and Ni, it decreases by  more than half at T > 800 K for Cr and Zn. For Zn, it could be attributed to its volatility while for Cr, it is possibly due to simulations using non ionized Cr or due to the formation of certain complexes of Cr-C which can escape out. The implantation fraction also decreases by half as relative velocities cross the 3000 $\mathrm{km}\,\mathrm{s}^{-1}$ threshold as transmission becomes dominant. Almost all the concentration of Zn found in SiC X grains can be attributed to implantation since it is volatile and any quantity of Zn co-condensed with SiC while grain condensation is highly likely to get evaporated. Maximum ppm concentrations of Zn predicted to be implanted in a 1$\micron$ grain are \textasciitilde1, 3 and 13 for 15, 20 and 25 M\textsubscript{\(\odot\)} models respectively. We also find that for a 5$\micron$ grain, even though less ions get implanted in it due to its larger size, lesser ions getting lost due to surface erosion is another factor that leads to a higher concentration of implantation for the larger grain. This happens because most ions are implanted into a 'penultimate' layer of the grain, at a depth below the outermost surface which gets eroded.
\item We also establish that transition ion implantation in the core of the grain is quite possible for a suitable range of velocities, thus increasing the chances of survival of transition elements in the grains. This is contradictory to the assumptions made so far about negligible impacts of ion implantation in the grain's core and encourages the hypothesis of smaller grains (rich in certain elements) to be embedded into bigger grains later on as sub-grains. Implantation is inhomogeneous and localized as opposed to condensation which is fairly homogeneous, however, it is more difficult than previously thought to observe this localization because all grain regions are within the reach of implantation, specially for smaller grains ($\la$ 1$\micron$). Hence, the measured concentrations which vary over 3 orders of magnitude reflect non negligible weightage from varied exposures to ion implantation scenarios in the SNe.
\item We find that the observed relative abundances (due to implantation and condensation) of Fe, Co and Ni can only be explained by considering mixing from innermost Ni and Si/S zones and are best matched against concentrations calculated for the 15 M\textsubscript{\(\odot\)} model. We find the mixtures we use to agree well with measured concentrations of Fe and Ni in SiC X grains from different mounts of the Murchison meteorite. Additionally, the S16 model confirms that such a mixing can also explain the isotopic abundance of Si in SiC X grains. This is in sync with 3D simulations and subsequent observations of SNRs which conclude that mixing happens as early as after few tens of seconds and ceases near \textasciitilde10$^{5-6}$ seconds. We work with two sets of mixtures from Ni, Si/S zone, He/C and He/N zones and follow two scenarios (for each mixture) for the role of $^{56}$Ni in the implantation of $^{56}$Fe owing to its inhomogeneous distribution in the form of clumps. Mixing from Si/S zone plays the most important role in calculating concentrations of Ni and Co. Additionally, mixing from He/C zones cannot be as high as 5\% or as low as < 0.5\% in order to explain the measured concentrations. Concentration of other two elements - Cr and Zn, is relatively lower (of the order of 1 and 0.1 ppm for a 15 M\textsubscript{\(\odot\)} model, 10 and 0.3 ppm  for a 20 M\textsubscript{\(\odot\)} model and 40 and 3 ppm for a 25 M\textsubscript{\(\odot\)} model). 
\item For grains where measured concentrations of Fe and Ni are $\ga$ 300 ppm, implantation fraction is $\la$ 0.25 and condensation dominates implantation whereas for other grains, implantation fraction can reach as high as \textasciitilde0.6. Implantation fraction of Ni is more than that of Fe possibly due to Ni being more volatile and hence having higher chances of evaporation after condensation. 
\end{enumerate}

Free expansion phase is the time period of maximum activity in the new born expanding ejecta, however, quantity of implantation beyond this period (where temperature-distance equations become highly non linear) must be investigated to account for changes in the grain structure which may dominate those set by this phase. Although this may be less likely since the grains might be able to develop protective layers of ice/organics or be embedded into larger grains, not to mention the decrement in density of particles due to volumetric expansion of the ejecta which will lead to lower probability of interactions of grains with ions. In any case, studies on galactic chemical evolution should help solve these mysteries.

\acknowledgments
\textit{Acknowledgments}
We are grateful to the anonymous referee whose comments helped improve this work. We acknowledge the utilization of Vikram 100 HPC Supercomputing Facility at PRL, Ahmedabad for computations related to SDTrimSP. We are indebted to Andreas Mutzke for his guidance while using SDTrimSP, Tuguldur Sukhbold for discussions on nucleosynthesis yields and Arkaprabha Sarangi for discussions on SiC condensation.

P.S. would like to thank Sanjukta Dhar and Akarsh Relhan for running simulations on SDTrimSP. The project was funded wide grant SERB-WE (964), Science and Engineering Research Board, Govt. of India.

\facilities{Vikram100 HPC} 
\software{SDTrimSP \citep{2011srim...74.1015M}, TRIM \citep{2002srim...74.1015Z,2010NIMPB.268.1818Z}}

\bibliographystyle{yahapj}
\bibliography{references}

\appendix
\section{Geometrical Considerations for ion Transmission through the grain}
\label{s:append}
Here, we describe the approximate geometrical model to calculate the fraction of ions transmitted through a spherical grain. Simulations in TRIM and SDTrimSP assume planar target surfaces whereas we assume the grains to be spherical. This model backtraces the linear trajectory of a transmitted ion to a sphere inside the planar surface and finds the \lq extra\rq\, length such a transmitted ion has to travel in the planar surface as compared to a spherical one (see section \ref{s:implan} for a discussion on the validity of linear trajectories). We perform this calculation of the extra lengths traversed for every ion which gets transmitted and a weight equal to the ratio of extra length by total length traversed is assigned to each transmitted ion. The weights are calculated separately for each dimension and then multiplied. For example, if the weight (extra length) for an ion transmitted in the simulation comes as 1.05, it implies that 1.05 ions would have transmitted had the surface been spherical.

\begin{figure*}
\centering
\figurenum{A1}
\label{fig:append}
\includegraphics[width=0.4\linewidth]{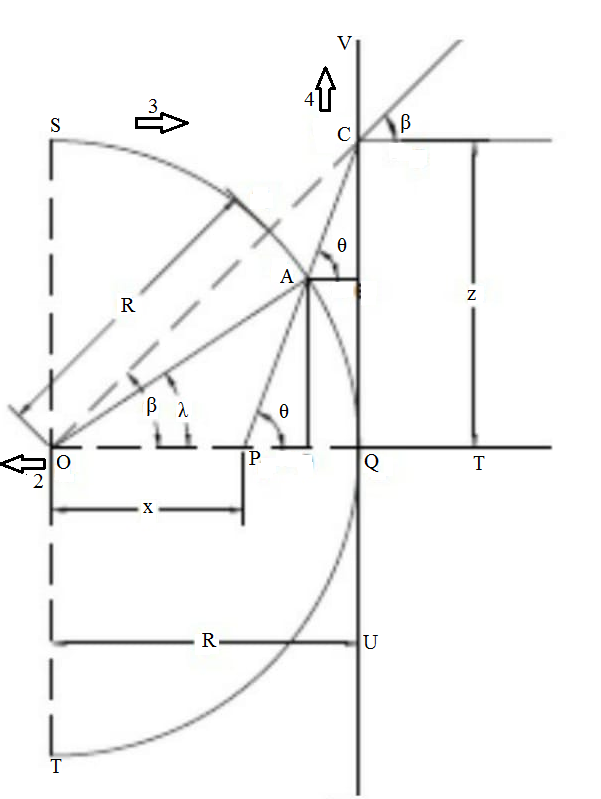}
\caption{Figure showing the first case as discussed in \ref{s:append} on the geometric setup to find the fraction of incident ions transmitted through a spherical grain. Arrow 2 corresponds to the second case, when the trajectory of transmitted ion backtraces it to a point on the other side (LHS) of the center (\textit{i.e.,} point P lies on the LHS of center O). Arrow 3 depicts the third case where a highly oblique incident ion is transmitted (in the direction S-A-T) whereas arrow 4 portrays the fourth case where the ion transmitted cannot be traced back to the grain at all.}
\end{figure*}

Let us call the length $\overline{AC}$ = $S_y$ (for y direction) and $S_z$ (for z direction). A total of four cases are developed for each axes such that each case in the region where y or z > 0 (called positive cases) has a corresponding case in the region where y or z < 0 (called negative cases). We note that $\theta \leq 135^{\circ}$ for positive cases otherwise the ion path would not trace back to the sphere. Similarly, $45\,<\,\theta\,<\,180$ for negative cases. The y and z values have been limited to $\pm$R. We only discuss the first case as illustrated in Figure \ref{fig:append}. The other three cases follow suite and can be worked upon in a straightforward manner. For this case, $0\,<\,\theta\,<\,90$, $\theta\,>\,\beta$, z (or y) $\leq$ R. Following the notations mentioned in Figure \ref{fig:append} and using cosine angle formula, following equations can be derived for $\Delta$OAC:

\begin{equation}
\angle OCA = \theta-\beta
\end{equation}

\begin{equation}
\mathit{\cos(\theta-\beta) = \frac{\overline{AC}^2+\overline{OC}^2-\overline{OA}^2}{2\,\overline{AC}\cdot\overline{OC}}}  
\end{equation}

Since, $\overline{OC} = \sqrt{R^2+z^2}$ and $\overline{OA} = R$,

\begin{equation}
\mathit{\cos(\theta-\beta) = \frac{\overline{AC}^2+z^2}{2\,\overline{AC}\cdot\sqrt{R^2+z^2}}}  
\end{equation}

\begin{equation}
\mathit{\overline{AC} = \cos(\theta-\beta)\sqrt{R^2+z^2}\pm\sqrt{(R^2+z^2)\,\cos^2(\theta-\beta)-z^2}}   \end{equation}

The positive solution is to be discarded since $\beta > 0$ for our model. Thus,

\begin{equation}
\mathit{\overline{AC} = \cos(\theta-\beta)\sqrt{R^2+z^2}-\sqrt{(R^2+z^2)\,\cos^2(\theta-\beta)-z^2}}   \end{equation}

where, $\overline{AC}$ is the extra length. Then, the weight is given by:

\begin{equation}
\mathit{W=\frac{L_{py}}{L_{py}-S_y}\frac{L_{pz}}{L_{pz}-S_z}}   \end{equation}

	where, $L_{pi}$ is the length traversed in the planar target, measured from the center in the $i^{th}$ direction.

There are a few cases (specially at very oblique angles of incidence) wherein the particle gets transmitted owing to its crossing the boundary of the target in only y or z direction. In such cases, the weights are computed as:

\begin{equation}
\mathit{W=\frac{L_{p\,(y,z)}}{L_{p\,(y,z)}-S_{\,(y,z)}}}
\end{equation}

where, the subscript $(y,z)$ implies either $y$ or $z$.

\end{document}